\documentclass{ws-rv9x6}
\usepackage{ws-rv-van}             

\def\to{\rightarrow}

\def\bi{\begin{itemize}}
 \def\ei{\end{itemize}}

\def\c1p{C1^\prime}

\def\ts{\tilde s}

\def\td{\tilde d}

\def\tst{\tilde t}
\def\ttau{\tilde \tau}

\def\tg{\tilde g}

\def\tw{\tilde\chi}

\def\tz{\tilde\chi^0}

\def\agt{\stackrel{>}{\sim}}
\def\be{\begin{equation}}  
\def\ee{\end{equation}}  
\def\bea{\begin{eqnarray}}  
\def\eea{\end{eqnarray}}  

\newcommand\prd[3]{{\it Phys.\ Rev.\ }{\bf D #1} (#2) #3}

\newcommand\plb[3]{{\it Phys.\ Lett.\ }{\bf B #1} (#2) #3}
\newcommand\jhep[3]{{\it J. High Energy Phys.\ }{\bf #1} (#2) #3}

\newcommand\zpc[3]{{\it Z.\ Physik }{\bf C #1} (#2) #3}
\newcommand\cpc[3]{{\it Comput.\ Phys.\ Commun.}{\bf #1} (#2) #3}

\newcommand{\hepph}[1]{hep-ph/#1}

\def\isajet{{\sc Isajet}}
\def\isasugra{{\sc Isasugra}}
\def\isasusy{{\sc Isasusy}}
\def\isatools{{\sc Isatools}}
\def\isared{{\sc IsaReD}}
\def\pythia{{\sc Pythia}}
\def\herwig{{\sc Herwig}}
\def\susygen{{\sc SUSYGeN}}
\def\sherpa{{\sc Sherpa}}
\def\softsusy{{\sc Softsusy}}
\def\suspect{{\sc SuSpect}}
\def\spheno{{\sc Spheno}}
\def\prospino{{\sc Prospino}}
\def\susyhit{{\sc SusyHIT}}
\def\hdecay{{\sc HDecay}}
\def\sdecay{{\sc SDecay}}
\def\feynhiggs{{\sc FeynHiggs}}
\def\cpsuperh{{\sc CPsuperH}}
\def\nmhdecay{{\sc NHMDecay}}
\def\comphep{{\sc CompHEP}}
\def\calchep{{\sc CalcHEP}}
\def\madgraph{{\sc MadGraph}}
\def\madevent{{\sc MadEvent}}
\def\grace{{\sc SUSY-Grace}}
\def\amegic{{\sc Amegic++}}
\def\lanhep{{\sc LanHEP}}
\def\helas{{\sc Helas}}
\def\omega{{\sc O'Mega}}
\def\whizard{{\sc Whizard}}
\def\grappa{{\sc Grappa}}
\def\darksusy{{\sc DarkSUSY}}
\def\micromegas{{\sc MicrOMEGAS}}
\def\sfitter{{\sc Sfitter}}
\def\fittino{{\sc Fittino}}

\begin{document}

\chapter*{Computational Tools for Supersymmetry Calculations\label{ch:baer}}

\author[H. Baer]{Howard Baer}

\address{Homer L. Dodge Department of Physics and Astronomy, \\
University of Oklahoma, Norman, OK 73019 USA\\
baer@nhn.ou.edu}

\begin{abstract}
I present a brief overview of a variety of computational tools for 
supersymmetry calculations, including: 
spectrum generators, 
cross section and branching fraction calculators, low energy constraints, 
general purpose event
generators, matrix element event generators, SUSY dark matter codes,
parameter extraction codes and Les Houches interface tools.
\end{abstract}

\body
\setcounter{chapter}{1}
\section{Introduction}\label{sec:intro}

The Standard Model (SM) of particle physics provides an excellent
description of almost all physical processes as measured in
terrestrial experiments, and is rightly regarded as the crowning achievement
of many decades of experimental and theoretical work in elementary particle
physics\cite{smreview}. 

As exciting as this is, it is also apparent that the SM cannot 
account for a wide assortment of astrophysical data, 
including neutrino oscillations,
the matter-anti-matter content of the universe, the presence of dark energy
and the presence of dark matter in the universe, 
and it doesn't include gravitation.
Even before these astrophysical anomalies became evident, it was apparent 
on theoretical grounds, mainly associated with quadratic divergences in 
the scalar (Higgs) sector, 
that the SM was to be regarded as an effective theory
valid only at the energy scale of $\sim 100$ GeV and below. 
At higher energies,
it seemed likely that some new physics must arise, which would be 
associated with the mechanism for electroweak symmetry breaking

While a vast array of physics theories beyond the SM have been proposed, 
the general class of theories including {\it weak scale supersymmetry}
seem to most naturally solve the theoretical ills of the SM, while at the
same time they receive support from a variety of 
precision experimental measurements\cite{wss}.
Most impressive is the measured values of three SM gauge couplings 
at the weak scale: when extrapolated to high energies using the 
renormalization group group equations, the gauge couplings very nearly 
meet at a point under supersymmetric standard model evolution, while they miss
badly under SM evolution\cite{gauge}. 
Gauge coupling unification suggests that
physics at scales $M_{GUT}\sim 2\times 10^{16}$ GeV is described by
a supersymmetric grand unified theory, and that below $M_{GUT}$, 
the correct effective field theory is the Minimal Supersymmetric Standard
Model (MSSM), or the MSSM plus gauge singlets 
(since gauge singlets don't affect the running of gauge couplings at one loop).

Supersymmetric models predict the existence of a whole new class
of matter states at or around the weak scale: the so-called super-partners.
Gluinos, charginos, neutralinos, squarks, sleptons plus 
additional Higgs scalars ($H$, $A$ and $H^\pm$) should all be present in
addition to the usual states of matter present in the SM.
In order to fully test the hypothesis of 
weak scale supersymmetry, it seems necessary to actually produce at
least some of the superpartners at high energy collider experiments,
and to measure many of their properties (mass, spin, coupling
strengths and mixing), in order to verify that any new physics signal
indeed comes from superpartner production. In addition, the properties
of the superpartners will be key to understanding the next level of
understanding in the laws of physics, perhaps opening windows to the
physics of grand unification and even string theory.

The key link between theoretical musings about various theories
of SUSY or other new physics, and the experimental observation of
particle tracks and calorimeter depositions in collider detectors is the 
{\it event generator program}. Given some theory of new physics, which
usually predicts the existence of new matter states or new interactions,
the event generator program allows us to compute 
how such a theory would manifest itself at
high energy colliding beam experiments. Thus, event generator programs function
as a sort of beacon, showing the way to finding new physics in a vast 
assortment of collider data.

Searches for new matter states
at the CERN LEP2 collider have found no firm new physics signals.
We thus conclude that the SM Higgs boson must have mass
$m_{H_{SM}}\agt 114$ GeV, while the charginos of supersymmetry must have mass
$m_{\tw_1}\agt 103.5$ GeV. The Fermilab Tevatron is probing sparticle masses
such as the gluino up to the 300 GeV level. 
The CERN LHC is a $pp$ collider which is just now beginning to explore
the energy regime where the SM breaks down, and where new physics ought 
to lie. LHC is expected to 
operate at energy scales $\sqrt{s}=7-14$ TeV; this ought to be sufficient 
to either produce some superpartners, or rule out most particle
physics models which include weak scale supersymmetry.

As we enter the LHC era, it is important to review the available calculational
tools which are available, that aid in connecting theory to experiment.
In this chapter, I present a brief overview of some of the publicly 
available tools. In Section \ref{sec:spec}, I examine the various 
\bi 
\item sparticle mass spectrum calculators,
\ei
and the {\it Les Houche Accord files} which
provide a handy interface between these and event generator programs.
Sec. \ref{sec:prodbf} lists some 
\bi
\item codes which calculate sparticle
production rates, decay widths and branching fractions. 
\ei
Sec. \ref{sec:evgen}
reviews 
\bi
\item event generators for SUSY processes, including 
\bi
\item multi-purpose
generators, complete with parton showers, hadronization and underlying events, 
and 
\item more specialized matrix element generators, which tend to focus on
specific reactions. 
\ei
\ei
The {\it Les Houche Event files} allow parton level
collider events to be easily passed into general purpose event 
generators so that showering, hadronization and underlying events 
can be included. 
Sec. \ref{sec:dm} lists 
\bi
\item codes relevant to supersymmetric dark matter calculations, 
\ei
while Sec. \ref{sec:fit}
examines 
\bi
\item codes designed to extract fundamental theory parameters from
sets of experimental measurements.
\ei
The supersymmetry parameter analysis (SPA) project seeks to develop a 
uniform set of conventions which would allow unambiguous extraction of 
high energy model parameters from various collider measurements of 
supersymmetric production and decay reactions.

I note here that this Chapter is an updated version of 
the 1997 version by H. Baer and S. Mrenna which appeared in the volume 
{\it Perspectives on Supersymmetry}, 
edited by G. Kane (World Scientific)\cite{old}.

\section{SUSY spectrum calculators}\label{sec:spec}

The first step in connecting supersymmetric theory to experiment is to begin
with a supersymmetric model, and calculate the expected spectrum of superpartner
and Higgs boson masses and couplings. 
The models we will be focusing on are 4-d supersymmetric quantum field 
theories with softly broken supersymmetry at the TeV scale. These models 
might be the low energy effective theory resulting from
some even more encompassing theory, 
such as superstring theory, or a particular
SUSY GUT model, or which may invoke some specific mechanism for supersymmetry
breaking.

The effective field theory is specified\cite{kane_review} by adopting 
1. the gauge symmetry, 2. the (super-) field content and 3. the Lagrangian.
In the case of supersymmetric theories, the Lagrangian is derived from
the more fundamental superpotential and K\"ahler potential, and for
non-renormalizable models, the gauge kinetic function. 
The effects of
SUSY breaking are encoded in the Lagrangian soft SUSY breaking (SSB) terms.
One must also specify the energy scale at which the effective theory 
and Lagrangian parameters are valid. Since collider experiments will be
testing physics at the weak scale $Q\sim 1$ TeV, while the 
Lagrangian parameters are frequently specified at much higher scales
({\it e.g.} $M_{GUT}$ or $M_P$),
the {\it renormalization group equations} (RGEs) must be used to
connect the disparate scales in the model.

Once the Lagrangian parameters are known at the weak scale, then the 
physical (s)particle masses must be identified, often by diagonalizing 
the relevant {\it mass matrices}. Higher order perturbative corrections to the
mass eigenstates-- at least at 1-loop-- are nowadays necessary to gain sufficient accuracy
in the predictions\cite{pbmz}.

Numerous researchers have developed private codes to calculate sparticle masses
given high scale model inputs. Here, we will focus only on {\it publicly available codes}, 
since these are available to the general user, and are frequently kept
up-to-date and user friendly. 
The first of the publicly available spectrum calculator codes to appear
was the \isasugra\cite{isasugra} subprogram of the event generator 
\isajet\cite{isajet}, in 1994. 
This was followed by \suspect\cite{suspect} (1997), 
\softsusy\cite{softsusy} (2002) and \spheno\cite{spheno} (2003).

\subsection{Isasusy, Isasugra and Isajet}

\isasusy\ is a subprogram of the \isajet\ event generator which calculates
sparticle mass spectra given a set of 24 SSB input parameters at the weak scale.
The program includes full 1-loop corrections to all sparticle masses. 
For Higgs masses and couplings,
the full 1-loop effective potential is minimized at an optimized scale choice which accounts for
leading 2-loop effects\cite{haber}. 
Yukawa couplings which are necessary for the loop calculations are evaluated
using simple SM running mass expressions.

The \isasugra\cite{isasugra} program starts with models defined at a much higher mass scale ({\it e.g.} $Q=M_{GUT}$),
and calculates the weak scale SUSY parameters via the full set of 2-loop RGEs\cite{mv}. 
An iterative approach to solving the RGEs
is employed, since weak scale threshold corrections which depend on the entire SUSY mass spectrum are included.
Once convergence is achieved, then the complete set of 1-loop corrected sparticle and Higgs masses are
computed as in \isasusy\ . Since \isasugra\ employs full 2-loop running of gauge and Yukawa couplings including
threshold corrections-- while \isasusy\ does not-- 
the sparticle masses will differ between \isasusy\ and \isasugra\ even for the same weak scale parameter
inputs. 

A listing of pre-programmed \isasugra\ models include the following:
\bi
\item mSUGRA (or CMSSM) model: 4 parameters plus a sign ($m_0,\ m_{1/2},\ A_0,\ \tan\beta ,\ sign(\mu )$),
\item minimal gauge-mediated SUSY breaking (mGMSB, 4 param's plus sign plus $C_{grav}$) and non-minimal GMSB,
\item non-universal supergravity (19 param's plus sign)
\bi
\item SSB terms can be assigned at any intermediate scale $M_{weak}<Q<M_{GUT}$,
\item non-universal Higgs model with weak scale $\mu$ and $m_A$ inputs in lieu of $m_{H_u}^2$ and $m_{H_d}^2$,
\ei
\item mSUGRA or NUSUGRA plus right-hand neutrino (RHN),
\item minimal and non-minimal anomaly mediation (AMSB),
\item mixed moduli-AMSB (mirage mediation),
\item hypercharged AMSB.
\ei

The related program \verb|RGEFLAV| computes the complete flavor matrix structure of the SSB terms and Yukawa
couplings, including $CP$-violating effects\cite{boxtata}.
The webpage is located at \verb|http://www.nhn.ou.edu/~isajet/|.

\subsection{Suspect}

\suspect\cite{suspect} runs the 2-loop MSSM RGEs to determine weak scale SUSY parameters in the mSUGRA, GMSB and AMSB
models, and in the pMSSM (a more general MSSM model). One-loop sparticle mass corrections are included.
Some two loop corrections to Higgs masses are included.
The webpage is located at \verb|http://www.lpta.univ-montp2.fr/users/kneur/Suspect/|.

\subsection{SoftSUSY}

\softsusy\cite{softsusy} is a $C++$ code that calculates 2-loop MSSM RGEs to determine weak scale SUSY parameters 
in the mSUGRA, mGMSB and mAMSB models, and in the general MSSM. 
$R$-parity violating effects are possible.
One-loop sparticle mass corrections are included.
Some two loop corrections to Higgs masses are included. \softsusy\ calculates the complete flavor matrix structure of
the MSSM soft terms and Yukawa couplings.
The webpage is located at \verb|http://projects.hepforge.org/softsusy/|.

\subsection{Spheno} 

\spheno\cite{spheno} is a Fortran 90 code that calculates 2-loop MSSM RGEs to determine weak scale SUSY parameters 
in the mSUGRA, mGMSB and mAMSB models, and in the general MSSM. One-loop sparticle mass corrections are included.
Some two loop corrections to Higgs masses are included. 
The webpage is located at \verb|http://www.physik.uni-wuerzburg.de/~porod/SPheno.html|.

\subsection{Les Houches Accord (LHA) files}

A standard input/output file under the name of Les Houches Accord (LHA) files has been created.
All of the above codes can create LHA output files. The advantage of LHA output files is that
various event generator and dark matter codes (see below) can use these as {\it inputs} for generating
collider events or dark matter observables. 
The specific form of the LHA files is presented in Ref. \cite{lha}.

In addition, the \isasugra\ and \isasusy\ codes output to a special \verb|IsaWIG| file, which
is created expressly for input of sparticle mass spectra and decay branching fractions to the
event generator \herwig\ . 

\subsection{Comparison of spectra generator codes}

Several papers have been written comparing the SUSY spectra codes\cite{kraml}, although these tend to
be all dated material, as the codes are continually being updated and debugged.
While many features of these codes are similar, and so agreement between spectra tends to be good
in generic parameter space regions, there are some differences as well. In particular, the codes
\suspect\ , \softsusy\ and \spheno\ all adopt a {\it sharp cut-off scale} between the MSSM and SM
effective theories. Allowance for the sharp cut-off is compensated for by $\log$ terms in the
1-loop sparticle and Higgs boson mass corrections. The \isasugra\ code instead 
adopts a ``tower of effective theories'' approach, and incorporates
threshold corrections in the 1-loop RGEs. Here, the beta-functions changes each time a sparticle mass 
threshold is passed over. 
One loop corrections to non-mixing sparticle masses are implemented at each sparticle's mass scale, so all 
$\log$s are minimized. Sparticles that mix have all their SUSY parameters evaluated at the 
$M_{SUSY}\equiv\sqrt{m_{\tst_L}m_{\tst_R}}$ 
scale due to a need for consistency amongst the various soft terms that enter the mass matrices\cite{bfkp}.
In this way, better accuracy is expected in cases where the sparticle mass spectra is spread across
a large energy range, as happens-- for instance-- in focus point or split 
SUSY, where scalars are at multi-TeV values or beyond,
whereas gauginos can be quite light. 

\section{Sparticle production and decay codes}\label{sec:prodbf}

\subsection{Production cross sections}

The multi-purpose event generators \isajet\ ,\pythia\ ,\herwig\ , \susygen\ 
and \sherpa\ all have a complete set of
tree-level SUSY particle production reactions encoded, and can be used to calculate tree-level 
sparticle production cross sections. In the case of \pythia\ or \herwig\ , 
the LHA files from spectra generators can be used as input to calculate these, 
or general SUSY parameter inputs are allowed.
\isajet\ does not allow LHA input since it has its own spectra generator. 
The \isajet\ code also calculates all sparticle and Higgs boson production reactions
for $e^+e^-$ colliders including variable beam polarization, and bremsstrahlung and beamstrahlung\cite{bmt}.
The \spheno\ code also calculates lowest order $e^+e^-\to SUSY$ cross sections. 

The code \prospino\cite{prospino} has been created to calculate all $2\to 2$ supersymmetric production 
cross sections at hadron colliders at both leading order (LO) and 
next-to-leading order (NLO) in QCD.
The current version of \prospino\ takes LHA files as its input format.

\subsection{Decay widths and branching fractions}

The programs \isasusy\ and \isasugra\ also calculate all sparticle and Higgs boson
$1\to 2$-body and $1\to 3$-body decay widths and branching fractions (BFs). 
These widths and BFs are output in \isajet\
standard output files, and are used internally for event generation. 
The chargino and neutralino branching fractions are sensitive to the parameter 
$\tan\beta$ in that at large $\tan\beta$, decays to third generation quarks and leptons
are enhanced relative to decays to first/second generation fermions.

The program \susyhit\cite{susyhit} is a relatively new release that combines \suspect\ with the 
branching fraction codes \sdecay\ and \hdecay\ to also generate a table of
sparticle and Higgs boson decay widths and branching fractions. Some of the
decay modes in \susyhit\ are calculated at NLO in QCD. 

The program \spheno\ also computes sparticle decay widths and branching fractions.

The branching fractions from \isajet\ , \susyhit\ and \spheno\ all seem to enjoy excellent
agreement with each other.
The branching fractions of all these codes can be input to event generators via the
LHA input/output files. Early versions of \herwig\ took branching fraction inputs from the
\verb|IsaWIG| output files.

Care must be taken in extracting branching fractions 
for neutralinos and charginos computed internally 
from \pythia\ in that they may not be valid at large $\tan\beta$ values $\agt 10$,
since Yukawa couplings, mixing effects, and decays through intermediate Higgs bosons
are neglected.

Some specialized codes are available for calculating decays modes of the SUSY Higgs bosons.
These include \feynhiggs\cite{feynhiggs}, which calculates MSSM Higgs boson masses at two-loop level, along with
branching fractions, \cpsuperh\cite{cpsuperh} which calculates Higgs boson branching fractions including CP-violating
parameters, and \nmhdecay\cite{nmhdecay}, which calculates Higgs boson masses and branching fractions in the
next-to-minimal supersymmetric Standard Model (NMSSM).

\section{Event generators}\label{sec:evgen}

Supersymmetric models can be used to calculate sparticle masses and mixings, which in turn 
allow for a prediction of various sparticle production rates and 
decay widths into final states containing quarks,
leptons, photons, gluons (and LSPs in $R$-parity conserving models). 
However, quarks and gluons
are never directly measured in any collider detector. Instead, 
detectors measure tracks of (quasi-)stable charged particles and their
momenta as they bend in a magnetic field. They also measure energy deposited 
in calorimeter cells by hadrons, charged leptons and photons.
There is thus a gap between the predictions of supersymmetric models
in terms of final states involving quarks, gluons, leptons and photons, with  
what is actually detected in the experimental apparatus. This gap is
bridged by event generator computer programs. Once a collider
type and supersymmetric model is specified,
the event generator program can produce a complete 
simulation of the sorts of scattering events that are to be expected.
The final state of any simulated scattering event is composed of 
a listing of electrons, muons, photons
and the long-lived hadrons (pions, kaons, nucleons etc.) and their
associated 4-vectors that may be measured in a collider experiment. 

The underlying idea of SUSY event generator programs
is that for a specified collider type ($e^+e^-$,
$pp$, $p\bar{p}$, $\cdots$) and center of mass energy, the event generator
will, for any set of model parameters,
generate various sparticle pair production events in the 
ratio of their production cross sections, and with distributions as given by their differential
cross sections. Moreover, the
produced sparticles will undergo a (possibly multi-step cascade) 
decay\cite{cascade} into a partonic final state, 
according to branching ratios as fixed by the model.
Finally, this
partonic final state is converted to one that is comprised of particles
that are detected in an experimental apparatus. By generating a large
number of ``SUSY events'' using these computer codes, the user can 
statistically simulate the various final states that are expected to be
produced within the framework of any particular model.  

Several general purpose event generator programs that incorporate SUSY are 
currently available, including \isajet\cite{isajet}, \pythia\cite{pythia}, 
\herwig\cite{herwig}, \susygen\cite{susygen} and \sherpa\cite{sherpa}.
These include usually just the leading order $2\to 2$ SUSY production processes.

In addition, 
specific $2\to n$ ($n\le 6$) SUSY reactions may be generated
by such programs as \comphep\ , \calchep\ , \madgraph\ , \grace\ , 
\amegic\ and \omega\ . The output of these
latter programs can be interfaced with the \pythia\ or \herwig\
programs to yield complete scattering event simulations by generating output
in the {\it Les Houches Event file} (LHE) format. 
(Isajet generates LHE output, but does not accept LHE files as input, since
it includes its own mass and branching fraction generator).
Ideally, event generator programs should be flexible enough to
enable simulation of SUSY events from a variety of models such as
mSUGRA, GMSB, AMSB {\it etc.}. This is usually accomplished nowadays by
reading in the LHA model files into the event generators \pythia\ or \herwig\ .
Since \isajet\ does its own spectra calculation, it only outputs LHA files, but does
not accept them as input.

\begin{figure}
\centerline{\psfig{file=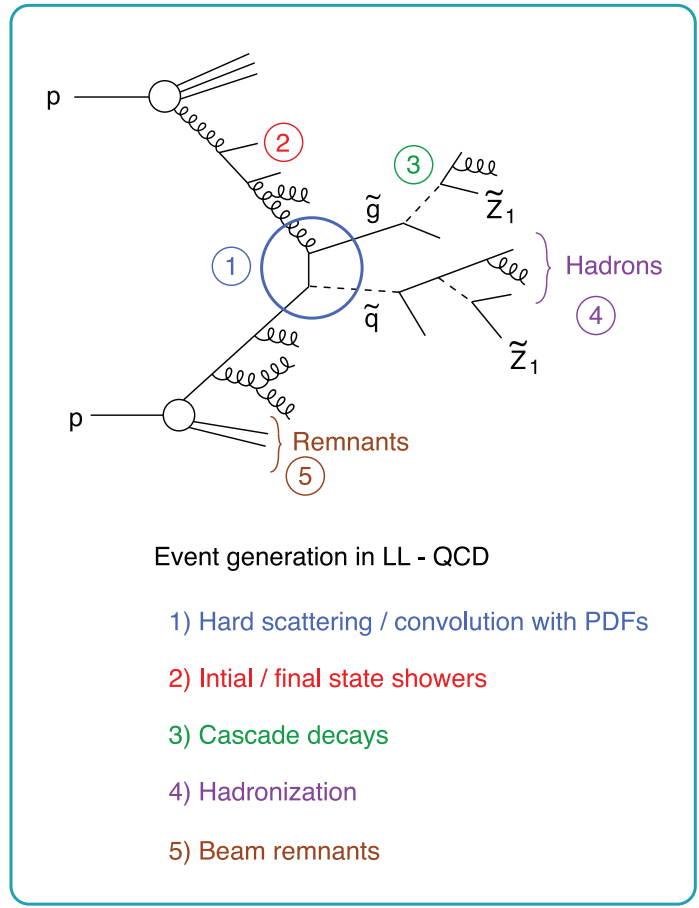,height=12cm}}
\caption[]{Steps in any event generation procedure.}
\label{fig:evgen}
\end{figure}

The simulation of hadron collider scattering events may be broken up into
several steps, as illustrated in Fig. \ref{fig:evgen}. The steps include:
\begin{itemize}
\item the perturbative calculation of the hard scattering subprocess
in the parton model, and convolution with parton distribution functions
(PDFs),
\item inclusion of sparticle cascade decays,
\item implementation of perturbative parton showers for initial and
final state colored particles, and for other colored particles
which may be produced as decay products of heavier objects,
\item implementation of a hadronization model which 
describes the formation of mesons and baryons from quarks and gluons. 
Also, unstable particles must be decayed to the (quasi-)stable daughters
that are ultimately detected in the apparatus, with rates and
distributions in  accord with their
measured or predicted values.
\item Finally, the debris from the colored remnants of the initial beams
must be modeled to obtain 
a valid description of physics in the forward regions of the collider 
detector. 
\end{itemize}
For $e^+e^-$ collider simulations, in addition we have to allow for 
the possibility of polarized initial beams, and beam-strahlung effects.

\subsection{Hard scattering}

The hard scattering and convolution with PDFs forms
the central calculation of event generator programs. 
The calculations are usually performed at lowest order in perturbation
theory, so that the hard scattering
is either a $2\to 2$ or $2\to 1$ scattering process. 
Matrix element generators are usually used for $2\to n$ processes, with $n\ge 3$.

For supersymmetric particle production at a high energy 
hadron collider such as the LHC, a 
large number of hard scattering subprocesses are likely to be
kinematically accessible. Each 
subprocess reaction must be convoluted with PDFs 
so that a total hadronic cross section for each reaction may be determined.
The $Q^2$-dependent PDFs commonly used are constructed to be solutions of the
Dokshitzer, Gribov, Lipatov, Altarelli, Parisi
(DGLAP) QCD evolution equations, which account for multiple
{\it collinear} emissions of quarks and gluons from the initial state
in the leading log approximation. 
As $Q^2$ increases, more gluons are radiated, so that the distributions
soften for large values of $x$, and correspondingly increase at small
$x$ values. 
Use of a running QCD coupling constant makes
the entire calculation valid at leading log level.

Once the total cross sections are evaluated for all the
allowed subprocesses, then
reactions may be selected probabilistically (with an assigned weight)
using a random number generator. This will yield sparticle events in the ratio
predicted by the particular model being simulated.

For sparticle production at $e^+e^-$ colliders, it may also be necessary
to convolute with {\it electron and positron} PDFs to incorporate bremsstrahlung and
beamstrahlung effects.
In addition, if beam polarization is used, then each
subprocess cross section will depend on beam polarization
parameters as well.

\subsection{Parton showers}

For reactions
occurring at both hadron and lepton colliders, to obtain a realistic
portrait of supersymmetric (or Standard Model) events,
it is necessary to account for multiple {\it non-collinear} 
QCD radiation effects. The evaluation of the cross section
using matrix elements for multi-parton final states is prohibitively
difficult. Instead, 
these multiple emissions are approximately included
in an event simulation via a parton shower (PS) algorithm.
They give rise to effects such as jet broadening, radiation in the forward regions
and energy flow into detector regions that are not described by
calculations with only a limited number of final state partons.

In leading log approximation (LLA), the cross section for {\it single} gluon
emission from a quark line is given by
\be
d\sigma =\sigma_0 \frac{\alpha_s}{2\pi}\frac{dt}{t}P_{qq}(z)dz ,
\label{eq:singleemm}
\ee
where $\sigma_0$ is the overall hard scattering cross section, 
$t$ is the intermediate state virtual quark mass, and 
$P_{qq}(z)= \frac{4}{3}\left(\frac{1+z^2}{1-z}\right)$
coincides with the Altarelli-Parisi splitting function for $q'\to qg$ for
the fractional momentum of the final quark
$z\equiv\frac{|\vec{p}_q|}{|\vec{p}_{q'}|}< 1$. 
Interference between various
{\it multiple} gluon emission Feynman graphs, where the gluons are ordered
differently, is a subleading effect. 
Thus, Eq.~(\ref{eq:singleemm}) can be applied successively, and gives a factorized
probability for each gluon emission. The idea behind the PS
algorithm is then to use these approximate emission
probabilities (which are exact in the collinear limit),
along with exact (non-collinear) kinematics to construct 
a program which describes multiple non-collinear parton emissions.
Notice, however,  that the cross section (\ref{eq:singleemm}) is singular
as $t\to 0$ and as $z\to 1$, {\it i.e.} in the regime of 
collinear and also soft gluon emission.
These singularities can be regulated by introducing physically appropriate
cut-offs. 
A cutoff on the value of $|t|$ of order $|t_c|\sim 1$ GeV 
corresponds to the scale
below which QCD perturbation theory is no longer valid. A cutoff
on $z$ is also necessary, and physically corresponds to the limit
beyond which the gluon is too soft to be resolved.

The PS algorithms available vary in their degree of sophistication.
The simplest algorithm was created by Fox and Wolfram in 1979\cite{fw}. 
Their method was improved to account for interference effects 
in the angle-ordered algorithm of Marchesini and Webber\cite{mw}.
In addition, parton emission from heavy particles results in a dead-cone
effect, where emissions in the direction of the heavy particle are 
suppressed. Furthermore, it is possible to include spin correlations
in the PS algorithm.

PS algorithms are also applied to the initial state partons. In this 
case, a backwards shower algorithm is most efficient, which develops
the emissions from the hard scattering backwards in time towards
the initial state. The backward shower algorithm developed
by Sj\"ostrand\cite{back} makes use of the PDFs evaluated at different energy scales 
to calculate the initial state parton emission probabilities.

\subsection{Cascade decays}

Not only are there many reactions available via which
SUSY particles may be produced at colliders, but once produced, 
there exist many ways in which superparticles may decay.
For the next-to-lightest SUSY particle (NLSP), there may be 
only one or at most a few ways to decay to the LSP. 
Thus, for a collider such as LEP or even the Fermilab Tevatron, where
only the lightest sparticles will have significant
production rates, we might 
expect that their associated decay patterns will be relatively simple.
However, the number of possible final states increases rapidly if
squarks and gluinos that can decay into the heavier charginos and
neutralinos are accessible, and
the book-keeping becomes correspondingly more complicated.  Indeed, at
the CERN LHC, where the massive strongly interacting sparticles such as
gluinos and squarks are expected to be produced at large rates, sparticle
cascade decay patterns can be very complex.
As a rough estimate, of order $10^3$ subprocess cross sections may be active at
LHC energies, with of order 10 decay modes for each sparticle. Naively, this would give of
order $10^5$ $2\to n$-body subprocesses that would need to be calculated.

Monte Carlo event generators immensely facilitate the analysis of
signals from such complex cascade decays, especially in the case
where no single decay chain dominates.  An event generator can select
different cascade decay branches by generating a random number which
picks out a particular decay choice, with a weight proportional to the
corresponding branching fraction, at each step of the cascade decay.
Quarks and gluons produced as the end products
of cascade decays will shower off
still more quarks and gluons, with probabilities determined by the PS
algorithm.  

The procedure that we have just described is exact for cascade decays
of spinless particles into two other spinless particles at each step in
the cascade. This is because the squared matrix element is just a
constant, and there are no spin correlations possible. This is not true
in general and
in many cases, it can be very important to include the decay matrix element
and/or spin correlations in the calculation of cascade decays of
sparticles. A general method for incorporating spin correlations based
on density matrices has been put forth by Richardson\cite{spincor}, 
and incorporated into \herwig\ .

Spin correlation effects are especially important for precision
measurements at $e^+e^-$ linear colliders. 
While retaining spin correlations may be less crucial in many
situations 
at a hadron collider, this is not always the case. 
For instance, relativistic $\tau^-$ leptons produced from $W$
decay are always left-handed, while those produced from a charged
Higgs decay are always right-handed. 
Likewise, the polarization of the taus from $\ttau_1$ decays depends on
the stau mixing angle. Since the undetectable 
energy carried off by $\nu_{\tau}$
from tau decay
depends sensitively on the parent tau helicity, it is necessary to include
effects of tau polarization in any consideration involving  the energy
of ``tau jets''.
By evaluating the mean polarization of taus
in any particular process, these effects can be
incorporated, at least on average, into  event generator programs
such as \isajet\ . 
Of course, such a procedure would not include correlations between 
decay products of two taus produced in the same reaction.

Another aspect is to include appropriately the complete 3-body decay
matrix elements. While some programs merely use a flat phase space
distribution, \isajet\ and \herwig\ include pre-programmed exact decay
matrix elements.

\subsection{Models of hadronization}

Once sparticles
have been produced and have decayed through their cascades,
and parton showers have been evolved up to the point where
the partons have virtuality smaller than $\sim 1$~GeV$^2$, 
the partons must be converted  to hadrons. This is a non-perturbative process,
and one must appeal to phenomenological models for its description.
The independent hadronization (IH) model of Field and Feynman\cite{ff}
is the simplest such model to implement. In this picture, 
a new quark anti-quark pair $q_1 \bar{q}_1$ can be created in the
color field of the parent quark $q_0$.
Then the $q_0\bar{q}_1$ pair can turn into a meson 
with a longitudinal momentum fraction described by a phenomenological
function, with the remainder of the longitudinal momentum carried
by the quark $q_1$. This process is repeated by the creation of
a $q_2\bar{q}_2$ pair in the color field of $q_1$,
and so on down the line to $q_n\bar{q}_n$. A host of mesons are 
thus produced, and decayed to the quasi-stable $\pi$, $K, \cdots $ mesons
according to their experimental properties. 
The final residual quark $q_n$ will have very little energy, and can be
discarded without significantly affecting jet physics.
Finally, a small transverse momentum 
can be added according to a pre-assigned Gaussian probability distribution
to obtain a better description of the data.
Quark fragmentation into
baryons is also possible by creation of 
diquark pairs in its color field, and can be incorporated.
The IH scheme, with many parameters tuned to fit the data, 
will thus describe the bulk features of
hadronization needed for event simulation programs.

The string model
of hadronization developed at Lund\cite{lund} is a more sophisticated
model than IH, which treats hadron production
as a universal process independent of the environment of the fragmenting 
quark. In the string model, a produced quark-antiquark pair is
assumed to be connected by a color flux tube or string. As the 
quark-antiquark
pair moves apart, more and more energy is stored in the string until
it is energetically favorable for 
the string to break, 
creating a new quark-antiquark pair. Gluons are regarded as kinks 
in the string. The string model correctly
accounts for color flow in the hadronization
process, as opposed to the IH model. In $e^+e^-\to q\bar{q}g$ (3-jet)
events, the string model predicts fewer produced hadrons in the regions
between jets than the IH model, in accord with observation.

A third scheme for hadronization is known as the cluster hadronization
model\cite{cluster}. In this case, color flow is still accounted for, but
quarks and antiquarks that are nearby in phase space will form a cluster,
and will hadronize according to preassigned probabilities. This model
avoids non-locality problems associated with the string
hadronization model, where quarks and antiquarks separated by spacelike
distances can affect the hadronization process.

\subsection{Beam remnants}

Finally, at a hadron collider the
colored remnants of the nucleon that did not participate in the hard
scattering must be accounted for. These beam remnant effects produce
additional energy flow, especially in the far forward regions of the
detector. A variety of approaches are available to describe these
non-perturbative processes, including models involving Pomeron exchange
and multiple scatterings. In addition, the beam remnants must be
hadronized as well, and appear to require a different parametrization
from ``minimum bias'' events where there are only beam jets but no
hard scattering.

\subsection{Multi-purpose event generators}

Publicly available event generators 
for SUSY processes include,
\begin{itemize}
\item {\bf ISAJET:} (H. Baer, F. Paige, S. Protopopescu and X. Tata),\break
\verb|http://www.nhn.ou.edu/~isajet/|
\item {\bf PYTHIA:} (T. Sj\"ostrand, L. L\"onnblad and S. Mrenna),\break
\verb|http://www.thep.lu.se/~torbjorn/Pythia.html|
\item {\bf HERWIG:} (G. Corcella, I. G. Knowles, G. Marchesini, S. Moretti,
K. Odagiri, P. Richardson, M. Seymour and B. R. Webber),\break
\verb|http://hepwww.rl.ac.uk/theory/seymour/herwig/|
\item {\bf SUSYGEN:} (N. Ghodbane, S. Katsanevas, P. Morawitz and E. Perez),
\break
\verb|http://lyoinfo.in2p3.fr/susygen/susygen3.html|
\item {\bf SHERPA:} (T. Gleisberg, S. H\"oche, F. Krauss, M. Sch\"onherr, S. Schumann,
F. Siegert and j. Winter)
\break
\verb|http://projects.hepforge.org/sherpa/dokuwiki/doku.php|
\end{itemize}

The event generator program \isajet\ was originally developed in the late
1970's to describe scattering events at the ill-fated ISABELLE $pp$
collider at Brookhaven National Laboratory. It was developed by F.~Paige
and S.~Protopopescu to generate SM and beyond scattering events at
hadron and $e^+e^-$ colliders. H. Baer and X. Tata, in collaboration 
with Paige and Protopopescu, 
developed \isajet\ to give a realistic
portrayal of SUSY scattering events.  \isajet\ uses the IH model for
hadronization, and the original Fox-Wolfram (Sj\"ostrand) PS shower
algorithm for final state (initial state) parton showers. It includes an
$n$-cut Pomeron model to describe beam-jet evolution.

The event generator \pythia\ was developed mainly by T. Sj\"ostrand 
in the early 1980s to implement the Lund string model for
event generation. \pythia\ uses the FW virtuality-ordered shower model, but 
with an angle-ordered veto. S.~Mrenna contributed the inclusion of
SUSY processes in \pythia\ .

The event generator \herwig\ was developed in the mid-1980s to
describe scattering events with angle-ordered parton showers, which
accounted for interference effects neglected in the FW
shower approach. \herwig\ implements a cluster hadronization model.
\herwig\ is notable in that it includes sparticle production and decay
spin correlations using density matrix techniques\cite{spincor}.

The program \susygen\ was developed by S.~Katsanevas and P. Morawitz to
generate $e^+e^-\to SUSY$ events for the LEP experiments. \susygen\
interfaces with \pythia\ for hadronization and showering. \susygen\ has
since been upgraded to also generate events for hadron colliders.

The program \sherpa\ was developed as a new generation event generator
in the $C++$ language.
It calculates subprocess reactions using \amegic\ . It includes its own
shower and cluster hadronization routines.

\subsection{Matrix element generators}

For generating various $2\to n$ scattering reactions using complete
matrix elements, a number of automated tree-level codes are available.

The code \comphep\ by E. Boos {\it et al.}\cite{comphep} is designed to take one 
directly from a Lagrangian to distributions. 
Feynman rules can be calculated using the \lanhep\ code\cite{lanhep}, 
and then \comphep\ will generate the squared matrix element by constructing the squared amplitude, 
taking traces, and storing the output as subroutines. \comphep\ also includes code for doing the phase space integration,
convolution with PDFs, and after integration, numerical output, or output in
terms of histograms.

The code \calchep\cite{calchep} by Pukhov, Belyaev and Christensen is very similar to \comphep\ , 
and was in fact created as a spin-off by some of the original authors 
of \comphep\ . 

The code \madgraph\ /\madevent\ was developed by Stelzer and Long 
and others\cite{madgraph}. It allows the user to input
initial and final state particles, and then generates all Feynman diagrams along with
a subroutine which evaluates the scattering amplitude as a complex number using the 
\helas\ helicity amplitude subroutines developed by Hagiwara and Murayama\cite{helas}.
Since \madgraph\ directly evaluates the amplitude, and not amplitude squared, computational sampling
of the squared matrix element should be faster than programs which evaluate traces over
gamma matrices. The latest versions of \madgraph\ , updated to \madevent\ , will convolute with PDFs
and perform phase space integration and evaluate distributions as well.
A number of models for BSM physics, including the MSSM , are available in 
\madgraph\ /\madevent\ .

The program \omega\ by Ohl, Reuter and Schwinn, also generates tree-level SM and MSSM 
amplitudes, and can work in concert with the \whizard\ program for event generation\cite{omega,whizard}.  

The program \grace\ by Tanaka, Kuroda, Kaneko, Jimbo and Kon also generates SM and MSSM amplitudes, and
generates scattering events in association with the \grappa\ program\cite{grace,grappa}.
\vspace{-.3cm}
\subsection{Les Houches Event (LHE) files}
\vspace{-.2cm}
A Les Houches Event (LHE) file format has been proposed\cite{lhe} which allows for a simple
communication between parton level event generators and all purpose generators
such as \pythia\ and \herwig\ . This is particularly useful when matrix element 
generators like \calchep\ or \madgraph\ are used, but the user needs a 
complete event output including parton showers, hadronization and underlying event
simulation. 

The LHE file is an \verb|ascii| file which includes lines pertaining to the generator
initialization. It then follows with a listing of partons (particle ID code), 
their associated 4-vectors and color flow information. The generators \pythia\ and \herwig\
then can read in these files, to add on showering, hadronization and underlying event.
A sample SUSY event in LHE format is listed below.
It lists a reaction with $sg\to\ts\tg$, with $\ts\to s\tz_1$ and $\tg\to\td \bar{d}$
and then $\td\to d\tz_1$. 
After listing a line of event characteristics, the event listing follows. 
The first column corresponds to particle ID, 2nd column to stability of particle, 3rd
and 4th columns list the source of the particle, 
5th and 6th columns relate to color flow, and 7th column
is the $x$-component of the energy-momentum four-vector.
The four-vector listing has been truncated to fit on the page. 
\begin{verbatim}
<event>
     10        2160        1.00000   0.768145E+06       ...
             3    -1     0     0  101    0  0.000000E+00  ...
            21    -1     0     0  102  101  0.000000E+00  ...
       2000003     2     1     2  103    0  0.220402E+03  ...
       1000021     2     1     2  102  103  -.220402E+03  ...
       1000022     1     3     0    0    0  0.778961E+02  ...
             3     1     3     0  103    0  0.142506E+03  ...
       2000002     2     4     0  102    0  -.185972E+03  ...
            -2     1     4     0    0  103  -.344299E+02  ...
       1000022     1     7     0    0    0  0.800434E+02  ...
             2     1     7     0  102    0  -.266016E+03  ...
</event>
\end{verbatim}

\section{Dark matter codes}\label{sec:dm}

In response to the increasing precision of data corresponding to the density of dark 
matter in the universe, several public codes have been developed which evaluate
key astrophysical observables in supersymmetric (and other) models. 

\subsection{DarkSUSY}

The \darksusy\ code, developed by Gondolo {\it et al.}\cite{darksusy}, evaluates the relic density of neutralino dark
matter in SUSY models. DarkSUSY computes all relevant neutralino annihilation and co-annihilation processes, and
solves the Boltzmann equation to output the current density of neutralino CDM. 
It accepts input files from Isajet/Isasugra or from LHA input files. DarkSUSY also calculates:
spin-independent and spin-dependent neutralino-nucleon scattering rates (direct WIMP detection), 
and indirect neutralino detection rates, such as: muon flux from neutralino annihilation in the core of
earth or sun, flux of $\gamma$ rays, $\bar{p}$s, $e^+$s and $\bar{d}$s from neutralino annihilation in
the galactic core or halo. The halo annihilation rates all depend on an assumed form for the galactic dark matter
density profile.

\subsection{Micromegas}

\micromegas\ was developed by Belanger {\it et al.}\cite{micromegas}, 
and also evaluates the neutralino relic density
due to all annihilation and co-annihilation processes. It also computes the WIMP relic density for a variety of other
non-SUSY models. It also outputs neutralino direct and indirect detection rates, 
$b\to s\gamma$ branching fraction and $(g-2)_\mu$.

\subsection{Isatools}

\isatools\ is part of the \isajet\ package. 
It includes a subroutine \isared\cite{isared} to evaluate the 
neutralino relic density, the direct neutralino detection rates via spin-independent and spin-dependent
scattering, the $b\to s\gamma$ branching fraction, $(g-2)_\mu^{SUSY}$, $BF(B_s\to \mu^+\mu^- )$ and
the thermally averaged neutralino annihilation cross section, which is key input to neutralino
halo annihilation calculations.


\section{Parameter fitting codes}\label{sec:fit}

If supersymmetry is indeed discovered at the Tevatron, LHC and/or a linear $e^+e^-$ collider, then
an exciting task will be to make precision measurements of all sparticle masses, spins, couplings
and mixings.
Once these are known, then, if the MSSM is indeed the correct effective theory all the way from
$M_{weak}$ to $M_{GUT}$, it is possible to map out the GUT scale values of the soft SUSY breaking parameters.
Once these are known, important information will be gained which will allow for the construction of
SUSY models at or beyond the GUT scale. Two such codes are available which accomplish this task:
\sfitter\cite{sfitter} and \fittino\cite{fittino}.

\section{SPA convention}

The supersymmetry parameter analysis (SPA) project\cite{spa} is an attempt to achieve co-ordination
between the various sparticle mass generation codes, event generators, relic density codes, and
parameter fitting codes, with a goal in mind to determine the fundamental SUSY Lagrangian.
In the SPA convention, all programs should input/ouput SUSY parameters in the $\overline{DR}$ scheme
at the $Q=1$ TeV scale. Once this benchmark is set, then all remaining calculations may proceed from this common
agreed upon point.

\section{Summary}

In the past decade, there has been an explosion of interest in supersymmetry
phenomenology. This is exhibited in part by the corresponding development 
of numerous computational tools to aid in supersymmetry calculations for
expected collider events and for dark matter observables.
Supersymmetry has certainly been an enduring theme in high energy physics.
Hopefully, at the dawn of the LHC era, 
we are on the verge of actual discovery of supersymmetry. In this case, 
many of these tools for SUSY will be put to good hard use, 
as the community analyzes the upcoming collider data. 

We expect that new tools 
for SUSY will emerge, which will be more focused on the 
new matter states that might appear. As an example, if SUSY is discovered, 
then the MSSM (or perhaps NMSSM) may become the new SM, and radiative 
corrections will have to be calculated for any remaining production and 
decay reactions, and in a form suitable for embedding in event generator
programs. 
The clues we find pertaining to dark matter will impact on all astrophysical
codes. In addition, 
new tools should also emerge that facilitate model building,
as the clues we expect to emerge from the data point the way to a new 
paradigm in physics beyond the Standard Model.

\section*{Acknowledgements} I thank Xerxes Tata for his comments on the 
manuscript, and Gordy Kane for urging me to write this updated chapter.

\end{document}